\documentclass[final,5p,times,twocolumn,number]{elsarticle}

\usepackage{amsmath,amssymb} 
\usepackage{graphicx}
\usepackage{xspace}
\usepackage{isotope}
\usepackage{orcidlink} 
\usepackage{lipsum}
\usepackage{siunitx}
\usepackage{upgreek}
\usepackage{booktabs}  
\usepackage{colortbl}
\usepackage{amsmath}
\usepackage[mathlines]{lineno} 
\usepackage[capitalise]{cleveref}
\usepackage{multirow}
\usepackage{microtype}
\usepackage{array}

\usetikzlibrary{shapes.geometric, arrows.meta, positioning, shadows, calc}
\tikzset{
  block/.style = {rectangle, rounded corners, draw, thick, minimum width=2.8cm, minimum height=1cm, align=center, fill=white, drop shadow},
  tool/.style  = {rectangle, draw, dashed, rounded corners, minimum width=2.8cm, minimum height=0.8cm, align=center, font=\small, fill=gray!5},
  arrow/.style = {-{Latex[length=3mm,width=2mm]}, thick},
  smallnode/.style = {font=\footnotesize, align=center}
}

\newcommand{\DDM}{\texttt{DirectDM}\xspace}
\newcommand{\DFF}{\texttt{DMFormFactor}\xspace}
\newcommand{\diamx}{\texttt{DIAMX}\xspace}
\newcommand{\apt}{\texttt{Appletree}\xspace}

\newcommand{\XEn}{\textnormal{XENONnT}\xspace}

\newcommand{\LZ}{\textnormal{LZ}\xspace}
\newcommand{\px}{\textnormal{PandaX-4T}\xspace}

\newcommand{\ty}{tonne $\times$ year\xspace}

\begin{document}


\title{Dark Matter implications from the LZ, PandaX-4T and XENONnT Data
}

\author[inst1,inst2]{Haipeng An}
\ead{anhp@tsinghua.edu.cn}

\author[inst1,inst2]{Fei Gao}
\ead{feigao@tsinghua.edu.cn}

\author[inst3,inst4]{Jia Liu}
\ead{jialiu@pku.edu.cn}

\author[inst5]{Minghao Liu}
\ead{ml5107@columbia.edu}

\author[inst1,inst2]{Haoming Nie}
\ead{nhm20@mails.tsinghua.edu.cn}

\author[inst1,inst2]{Changlong Xu}
\ead{xuchanglong@mail.tsinghua.edu.cn}

\address[inst1]{Department of Physics, Tsinghua University, Beijing 100084, China}
\address[inst2]{Center for High Energy Physics, Tsinghua University, Beijing 100084, China}
\address[inst3]{School of Physics and State Key Laboratory of Nuclear Physics and Technology, Peking University, Beijing 100871, China}
\address[inst4]{Center for High Energy Physics, Peking University, Beijing 100871, China}
\address[inst5]{Physics Department, Columbia University, New York, NY 10027, USA}

\begin{abstract}
We investigate a possible dark matter origin of the high-energy nuclear-recoil-like events in data from liquid xenon time projection chamber experiments, including \LZ, \px, and \XEn, which cannot be explained by standard elastic spin-independent WIMP scattering. Using our unified \diamx framework, built on openly available data and likelihood models, we perform the first combined profile-likelihood fits to multiple WIMP-search datasets with a total exposure of approximately 8.8~\ty. We consider two broad classes of dark matter–nucleon interactions, involving either velocity-dependent cross sections or inelastic (endo- and exothermic) scattering, which can reproduce the observed high-energy recoil spectrum, reaching local significances up to $3.5\sigma$. We further quantify the impact of $^{124}$Xe double electron capture (DEC) backgrounds, finding that variations in the poorly known DEC charge yields can shift the inferred significances from a null-like result to $3.5\sigma$.
We further note that extending the same analysis to data from all three experiments with recoil energies up to $300~\mathrm{keV}$, when available, will provide a powerful test of the dark matter interpretation, since the $^{124}$Xe DEC background is expected to be negligible in this high-energy range.
\end{abstract}


\maketitle

\noindent \textbf{Introduction.}
Dark matter (DM) accounts for approximately one-fourth of the universe's total energy density today, yet its particle nature remains enigmatic. While various alternative models have been proposed, the weakly interacting massive particle (WIMP) remains the most compelling paradigm for DM due to its theoretical elegance and natural ability to explain the observed relic abundance through thermal freeze-out~\cite{HUT197785,Weinberg1977}. Experiments investigating the direct interaction between WIMPs and Standard Model particles have produced stringent limits on the elastic spin-independent WIMP-nucleon cross section. Among them, experiments searching for nuclear recoil (NR) signals in liquid xenon (LXe) time projection chambers (TPC) are leading for DM masses ranging from a few GeV to tens of TeV~\cite{LZ:2022lsv,LZ:2024zvo,XENON:2023cxc,XENON:2025vwd,PandaX:2024qfu}, thanks to their sensitivity to low-energy events, low background levels, and excellent discrimination power between the dominant electronic recoil (ER) background and the NR signal.

\begin{figure}[!ht]
\includegraphics[width=0.95\linewidth]{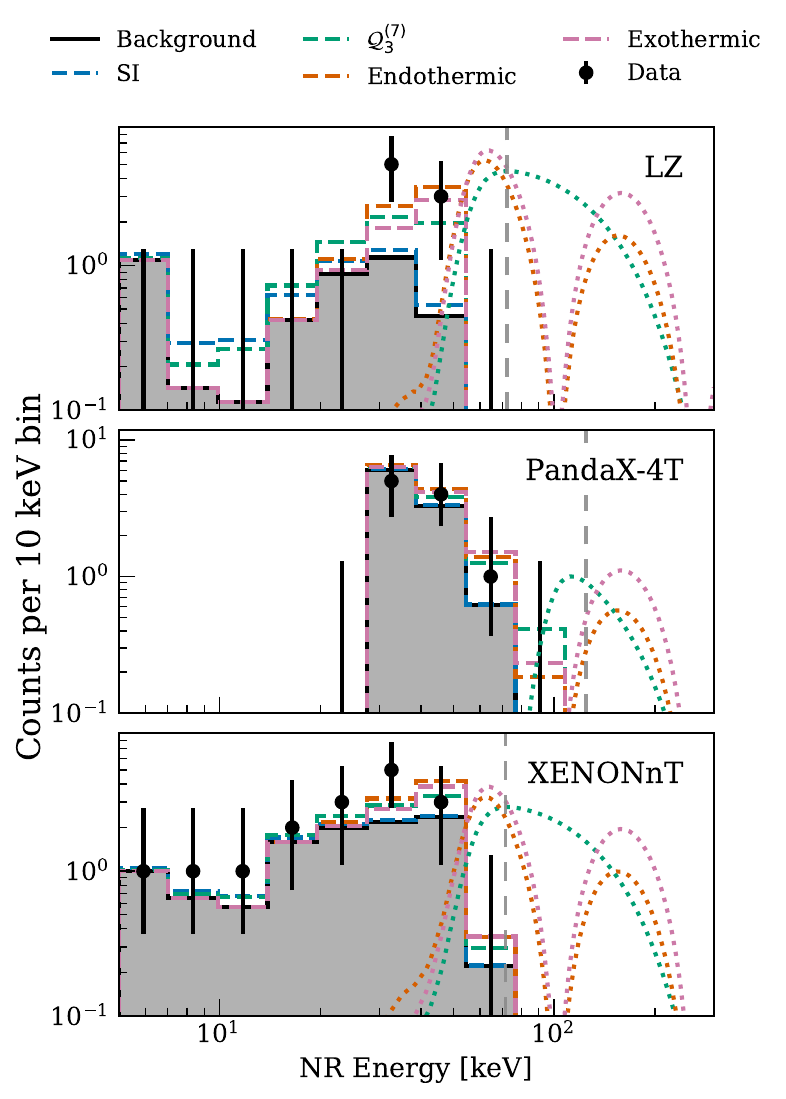}
    \caption{
    Event distributions in NR energy for \LZ (top), \px (middle), and \XEn (bottom). Data points are selected below or at the NR median and above the $2\sigma$ lower bound of the NR band in corrected (S1, S2); for \XEn we additionally require cS2 $>$ 400 PE, while for \px we require cS1 $>$ 30 PE to mitigate low-energy mismodeling. Error bars show 1$\sigma$ Feldman–Cousins intervals for signal+background counts. The black solid curve and the gray band give the predicted background under the same selection criteria, while the colored dashed curves show background plus signal for four DM benchmarks: spin-independent (SI) elastic scattering with $m_{\mathrm{DM}}=200~\mathrm{GeV}$ (fixed to the 90\% C.L. upper limit of the cross section reported in Ref.~\cite{LZ:2024zvo}), $\mathcal{Q}_3^{(7)}$ with $m_{\mathrm{DM}}=100~\mathrm{GeV}$, endothermic inelastic scattering with $\delta = 150~\mathrm{keV}$ and $m_{\mathrm{DM}} = 1~\mathrm{TeV}$, and exothermic inelastic scattering with $\delta = -200~\mathrm{keV}$ and $m_{\mathrm{DM}} = 1~\mathrm{TeV}$, the latter three evaluated at their combined \textnormal{XENONnT}-\textnormal{LZ} best-fit normalizations. Dotted lines show the predicted number of events outside the WIMP search region of interest (ROI), using the same normalization and extrapolated efficiencies but omitting the NR-band selections. Dashed vertical lines show the maximum NR energies where the efficiencies (including region-of-interest) of the corresponding experiments exceed 1\%.
    }
    \label{fig:nr_below_median}
\end{figure}

Although no convincing evidence of WIMP-nucleus elastic scattering has been found, various LXeTPC experiments have independently reported NR candidate events at energies higher than $20~\mathrm{keV}$. The XENON1T collaboration observed 14 events in the NR signal reference region with a background expectation of $7.36 \pm 0.61$ using 1.0 tonne$\times$year exposure~\cite{XENON:2018wimp}. In the more recent WIMP search datasets from LZ, PandaX-4T and XENONnT, excesses of NR-like events in their respective WIMP search regions were also observed, with a significant fraction lying above 20 keV, although independently their data are consistent with the background-only models~\cite{PandaX:2024qfu,XENON:2025vwd}. In this work, we examine the WIMP search datasets from \LZ~\cite{LZ:2022lsv,LZ:2024zvo}, \px~\cite{PandaX:2024qfu} and \XEn~\cite{XENON:2023cxc,XENON:2025vwd}, with total exposures of 4.2, 1.54 and 3.1 tonne$\times$year, respectively. As shown in Fig.~\ref{fig:nr_below_median}, we observe consistent NR-like events above $20\,\rm keV_{NR}$ in the data from all three experiments. The standard spin-independent WIMP–nucleus elastic scattering model cannot reproduce such spectral features due to its exponentially falling recoil spectrum.

The \LZ collaboration interpreted the NR-like events as originating from the two-neutrino double electron capture (2$\nu$ECEC, or DEC) of $^{124}$Xe,
\[
^{124}\mathrm{Xe} + 2e^- \to {}^{124}\mathrm{Te} + 2\nu_e ,
\]
a rare second-order weak process with a measured half-life of
$T_{1/2}^{2\nu\mathrm{ECEC}} \simeq 1.8\times10^{22}~\mathrm{yr}$~\cite{XENON:2019dti}.
Sub-dominant capture channels, such as double L-shell (LL) capture, deposit a relaxation energy of $10.01~\mathrm{keV}$ with a capture probability of $1.22\%$~\cite{LZ:2025hud}, leading to low-energy and highly localized atomic de-excitation cascades.
These events are expected to exhibit enhanced recombination and a reduced charge yield, allowing DEC backgrounds to leak into the nuclear-recoil (NR) region~\cite{LZ:2024zvo,Xu:2025ksk}. However, the exact charge yield is difficult to measure due to the rarity of the decay, and introducing a free parameter for it may mask potential signals in the NR region. In contrast, the \XEn collaboration did not incorporate this interpretation into their main result because their dataset failed to reject the model with a nominal DEC charge yield~\cite{XENON:2025vwd}. The \px collaboration included the LL-shell DEC component in their background model, but their analysis did not introduce a dedicated reduced-charge-yield treatment for this component~\cite{PandaX:2024qfu}.

In this Letter, we present the first combined profile-likelihood analysis of the high-energy NR-like events across LZ, PandaX-4T, and XENONnT. We show that standard elastic spin-independent WIMP scattering struggles to explain the observed spectral features, whereas specific velocity-dependent and inelastic (endo- and exothermic) scattering scenarios can provide a significantly better description. At the same time, we identify the poorly constrained charge yields of $^{124}$Xe double electron capture (DEC) as the dominant systematic uncertainty that strongly influences the apparent preference for dark matter signals. We further suggest that future analyses by \LZ, \px, and \XEn extending to higher recoil energies ($\gtrsim 100-300\,\rm keV_{NR}$) will be crucial to distinguish a potential dark matter signal from the DEC background. 
\\

\begin{table*}[t]
\centering
\footnotesize
\caption{Ratios of $Q_L$, $Q_{LM}$, and $Q_{LL}$ to $Q_\beta$ for the different DEC charge-yield scenarios and datasets. The label \textit{float} indicates that the ratio is treated as an independent free parameter in the combined fit. The last two columns show the maximum local signal significance and the corresponding best-fit parameters for each scenario.}
\label{tab:q_ratios}

\setlength{\tabcolsep}{6pt}
\renewcommand{\arraystretch}{1.4}

\begin{tabular}{c|c|c|c|p{1.5cm}p{1.cm}p{4cm}}
\toprule
\textbf{Case} & \textbf{Experiment} & $Q_{LM} / Q_\beta$ & $Q_{LL} / Q_\beta$ & Model & Sig. & Best-fit Para.~ ($\Lambda,~m_{\mathrm{DM}},~|\delta|$) \\
\midrule

I &
\begin{tabular}{c}
\LZ\\
\px\\
\XEn
\end{tabular}
&
\begin{tabular}{c}
0.88\\
0.88\\
1.00
\end{tabular}
&
\begin{tabular}{c}
0.88\\
0.88\\
1.00
\end{tabular}
&
\begin{tabular}{@{}l@{}}
$\mathcal{Q}_3^{(7)}$\\
$\mathcal{Q}_4^{(7)}$\\
Endothermic\\
Exothermic
\end{tabular}
&
\begin{tabular}{@{}l@{}}
$2.3\sigma$\\
$2.6\sigma$\\
$3.5\sigma$\\
$3.5\sigma$
\end{tabular}
&
\begin{tabular}{@{}l@{}}
(31.3~GeV, 79~GeV, --)\\
(2.24~GeV, 55~GeV, --)\\
(3.14~TeV, 60~GeV, 130~keV)\\
(93.9~TeV, 13~GeV, 455~keV)
\end{tabular}
\\
\midrule

II &
\begin{tabular}{c}
\LZ\\
\px\\
\XEn
\end{tabular}
&
\begin{tabular}{c}
0.88\\
0.88\\
1.00
\end{tabular}
&
\begin{tabular}{c}
\textit{float}\\
\textit{float}\\
1.00
\end{tabular}
&
\begin{tabular}{@{}l@{}}
$\mathcal{Q}_3^{(7)}$\\
$\mathcal{Q}_4^{(7)}$\\
Endothermic\\
Exothermic
\end{tabular}
&
\begin{tabular}{@{}l@{}}
$0.1\sigma$\\
$0.5\sigma$\\
$1.8\sigma$\\
$1.8\sigma$
\end{tabular}
&
\begin{tabular}{@{}l@{}}
(45.7~GeV, 79~GeV, --)\\
(2.64~GeV, 55~GeV, --)\\
(3.23~TeV, 56~GeV, 125~keV)\\
(113~TeV, 12~GeV, 425~keV)
\end{tabular}
\\
\midrule

III &
\begin{tabular}{c}
\LZ\\
\px\\
\XEn
\end{tabular}
&
\begin{tabular}{c}
0.88\\
0.88\\
\textit{float}
\end{tabular}
&
\begin{tabular}{c}
\textit{float}\\
\textit{float}\\
\textit{float}
\end{tabular}
&
\begin{tabular}{@{}l@{}}
$\mathcal{Q}_3^{(7)}$\\
$\mathcal{Q}_4^{(7)}$\\
Endothermic\\
Exothermic
\end{tabular}
&
\begin{tabular}{@{}l@{}}
$0$\\
$0$\\
$1.1\sigma$\\
$1.0\sigma$
\end{tabular}
&
\begin{tabular}{@{}l@{}}
(--, --, --)\\
(--, --, --)\\
(2.09~TeV, 54~GeV, 125~keV)\\
(132~TeV, 10~GeV, 495~keV)
\end{tabular}
\\

\bottomrule
\end{tabular}
\end{table*}

\noindent \textbf{Velocity-dependent and inelastic interactions}.\\
Motivated by the high-energy NR-like events and the limited understanding of the DEC charge yield, we investigate DM interpretations of the \LZ,\px and \XEn data.
Since these events in all experiments are above 20 $\rm keV_{NR}$, in this work, we consider two classes of DM models:
The first class involves velocity-dependent DM-nucleus elastic interactions, systematically described within the DM Effective Field Theory (DMEFT)~\cite{Cao:2009uw, An:2010kc, Fan:2010gt, Fitzpatrick:2012ix, Gluscevic:2015sqa, DeSimone:2016fbz, Liem:2016xpm, Bishara:2017pfq, Song:2023jqm}.
Two representative dimension seven operators entering our analysis are
\begin{equation}
\begin{aligned}
\mathcal{Q}_3^{(7)} &= \frac{1}{\Lambda^3}\frac{\alpha_s}{8\pi} (\bar{\chi}\chi)G^{a\mu\nu}\widetilde{G}^a_{\mu\nu},\\
\mathcal{Q}_4^{(7)} &= \frac{1}{\Lambda^3}\frac{\alpha_s}{8\pi} (\bar{\chi} i\gamma_5 \chi)G^{a\mu\nu}\widetilde{G}^a_{\mu\nu},
\end{aligned}
\end{equation}
Here, $\chi$ denotes a spin-1/2 Dirac fermion DM particle, $G_{\mu\nu}^a$ ($\widetilde{G}_{\mu\nu}^a$) are the gluon field-strength tensor and its dual, and $\Lambda$ denotes the cutoff scale. We perform a systematic scan over the operators in the Chiral EFT framework~\cite{Bishara:2017pfq}. Among all operators considered in the scan, we find that only a limited subset, in particular $\mathcal{Q}_3^{(7)}$ and $\mathcal{Q}_4^{(7)}$, can provide an acceptable description of the observed spectral features. In the nonrelativistic limit, their DM–nucleus scattering amplitudes scale as $q$ and $q^2$, respectively, where $q$ is the momentum transfer. Consequently, the corresponding recoil-energy spectra are suppressed at low $E_\mathrm{R}$.

These partonic operators are matched onto their nonrelativistic counterparts at leading order in chiral counting~\cite{Bishara:2017pfq}. The resulting scattering cross sections are expressed in terms of nuclear response functions, which are mainly obtained from theoretical nuclear-structure calculations and partially validated against experimental data~\cite{Fitzpatrick:2012ix}. The matching of partonic DMEFT operators to the nonrelativistic EFT is performed using the Mathematica package \DDM~\cite{Bishara:2017nnn}, and the resulting recoil spectra are then computed with \DFF~\cite{Anand_2014}. We assume a standard Maxwell–Boltzmann velocity distribution for the local dark matter halo; a detailed exploration of astrophysical uncertainties is beyond the scope of this work.

The second class corresponds to inelastic DM. In this class, the DM particle transitions to a slightly heavier or lighter state with a small mass splitting, $m'_{\rm DM} = m_{\rm DM} + \delta$, after scattering with a nucleon.
Both endothermic ($\delta > 0$)~\cite{Tucker-Smith:2001myb} and exothermic ($\delta < 0$)~\cite{Graham:2010ca,Batell:2009vb} scenarios are considered. In order to isolate the effect of inelastic scattering, we focus on the vector-vector spin-independent (SI) interaction $\frac{1}{\Lambda^2}\left(\bar{\chi} \gamma_\mu \chi'\right)\left(\bar{q} \gamma^\mu q\right)$. 
In the endothermic case, the minimum velocity required to produce a given recoil energy $E_R$ is
\begin{equation}
v_{\min }\left(E_{\mathrm{R}}\right)=\frac{1}{\sqrt{2 m_N E_{\mathrm{R}}}}\left|\frac{m_N E_{\mathrm{R}}}{\mu_N}+\delta\right|,
\label{eq:vmin}
\end{equation}
where $\mu_N$ is the reduced DM-nuclei mass.
The recoil spectra are computed following the same procedure as in the first class, and the rate spectra are obtained using the identical tool chain.
\\

\noindent \textbf{Data analysis}.
Our analysis includes six datasets: the 2022 and 2024 WIMP search datasets from \LZ (WS2022 and WS2024), the Run 0 and Run 1 datasets from \px, and the Science Run 0 and Science Run 1 datasets from \XEn (SR0 and SR1). For each dataset, we incorporate the full signal and background models of \LZ, \px, and \XEn into a unified statistical inference framework. In these experiments, particle interactions in LXe produce a prompt scintillation signal (S1) and a delayed electroluminescence signal (S2) from ionized electrons extracted into the gas phase. Each event is represented in the 2D plane of corrected S1 and S2, where the corrections remove spatial and temporal dependencies, and the ratio S2/S1 provides statistical discrimination between nuclear recoils (NR) and electronic recoils (ER) events. Two sets of units are used when reporting the size of the S1 and S2 signals, depending on whether the double-photoelectron (DPE) effect of the PMT is corrected from raw PMT signals. We denote quantities in photons detected (phd, DPE correction applied, used in \LZ) as (S1c, S2c), and quantities in photoelectrons (PE, DPE correction not applied, used in \px and \XEn) as (cS1, cS2). \px only uses the bottom PMTs for S2 signals, so their corrected S2 sizes are denoted as $\mathrm{cS2}_{\mathrm{b}}$. Backgrounds in both experiments fall into three broad categories: ER backgrounds, dominated by $\beta$ decays of $^{214}$Pb in the $^{222}$Rn chain and characterized by higher S2/S1 than NR events; NR backgrounds from radiogenic neutrons and CE$\nu$NS; and data-driven backgrounds, including accidental coincidences (AC) in both experiments and surface events in \XEn.

We use our \diamx framework~\cite{DIAMX} to model the event distributions of each background and signal component in corrected S1–S2 space. \diamx relies only on publicly available information and is designed to reproduce the official collaboration models as closely as possible. For simplicity, we omit CE$\nu$NS in \LZ and \XEn, neutrons in \LZ, and surface events in \px and \XEn SR1, as their contributions are negligible. The position- and time-dependent correction maps are not public and are not included in our analysis.

The ER and NR models for \LZ are based on a customized version of \texttt{NEST v2.4.0}~\cite{nest:paper,nest:software}, with parameters taken from the publicly released configuration files~\cite{LZ:2022lsv,LZ:2024zvo}; the published ER and NR efficiencies are used directly.
Data-driven backgrounds are more challenging to model due to limited public information. We assume a Gaussian-like profile for the AC background and a piecewise uniform profile for the surface background, with parameters chosen such that the published 68\% and 95\% contours are exactly reproduced.

For \px, the yield and detector models follow Ref.~\cite{PandaX:signal_response} and are implemented in the \apt~framework~\cite{xu_xenonntappletree_2025}. The yield parameters are refitted to the published light and charge yields for ER and NR events, respectively. The ER and NR detection efficiencies are modeled as the product of S1- and S2-related selection efficiencies, reconstruction efficiencies, region-of-interest efficiencies, and other analysis cuts. An electron-lifetime correction is further applied to improve the description of the ER and NR band distributions.

The ER and NR modeling \XEn is also based on the software \apt~\cite{xu_xenonntappletree_2025}. The microphysics (NEST) parameters, describing the light and charge yields of xenon, are taken from Ref.~\cite{XENON:analysis_paper_2} for both SR0 and SR1, and the detector parameters follow Refs.~\cite{XENON:2023cxc,XENON:2025vwd}, except that the electron gain $g_2$ in SR1 is adjusted to $15.8~\text{PE/e}^-$, instead of the published $(16.9 \pm 0.5)~\text{PE/e}^-$. This adjustment is introduced to account for the differences between our recast model and the internal XENONnT model, so that the published ER and NR band locations in the $\mathrm{cS1}-\mathrm{cS2}$ plane can be reproduced. An additional S1 reconstruction efficiency correction is applied to the \XEn efficiency following Ref.~\cite{XENON:analysis_paper_one}.

{
\setlength{\textfloatsep}{4pt}
\begin{figure*}[!t]
    \centering
   \includegraphics[
   width=\textwidth,
   height=0.65\textheight,
   keepaspectratio]{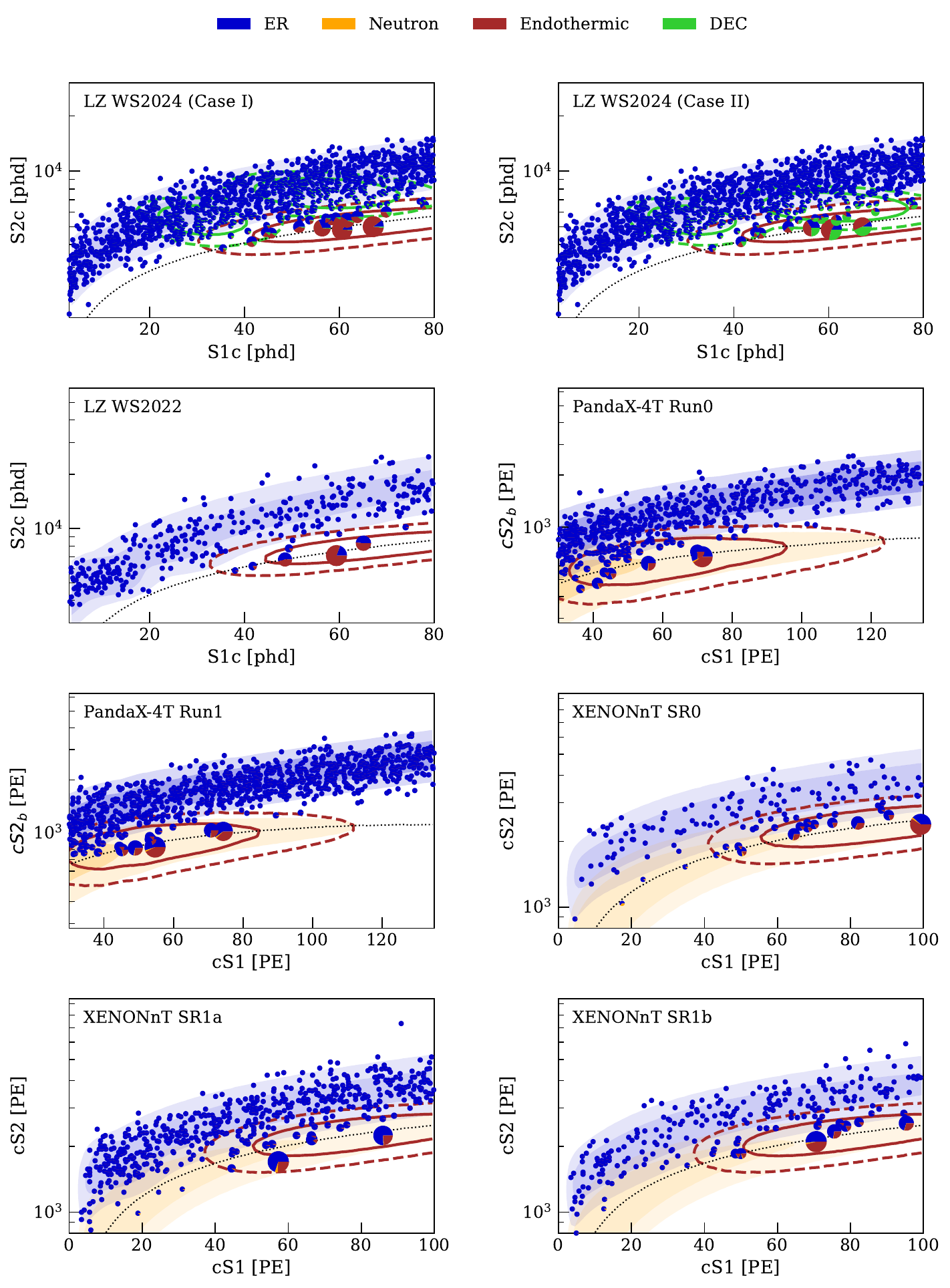}
    \caption{
    Corrected (S1, S2) distributions for all \LZ, \px, and \XEn datasets used in this analysis. 
    Each event is represented by a pie chart showing the fractional contributions of the local probability density from different components of the best-fit endothermic inelastic DM model at $\delta = 150~\mathrm{keV}$, $m_{\mathrm{DM}} = 100~\mathrm{GeV}$. The pie size is proportional to the local endothermic signal fraction. 
    Shaded regions indicate the $1\sigma$ and $2\sigma$ contours for ER (blue) and neutron (orange) backgrounds. 
    Red solid and dashed curves show the corresponding contours for the endothermic signal, and green curves are shown for LZ WS2024 for the DEC background under case I and case II treatment. Dotted lines show the NR median curves for each dataset and illustrate the data selection criteria used in Fig.~\ref{fig:nr_below_median}.
    }
    \label{fig:five_figures}
\end{figure*}
}

The event distributions generated by \diamx agree well with the published background contours, except for NR events at low energies in \XEn, and for both ER and NR events at low energies in \px. We therefore apply an additional selection of $\text{cS2} > 400$~PE to all \XEn datasets, and $\text{cS1} > 30$~PE to all \px datasets. While the additional \XEn selection has a negligible impact on the high-energy region, the \px cut removes events up to approximately $30~\text{keV}_{\mathrm{NR}}$, which overlaps with the parameter space of several dark matter models considered in this work. This cut thus constitutes a conservative treatment of this region in the present analysis. This discrepancy indicates that a refined understanding of the low-energy \px detector response is required.

The decomposition of background components, along with their rates and associated uncertainties, follows the corresponding publications. The recoil spectra of the neutron background for both \px and \XEn are taken from Ref.~\cite{XENON:sensitivity}, and the recoil spectra of the CE$\nu$NS background for \px from Ref.~\cite{PandaX:cevns}. Most $\beta$ decays are modeled with uniform spectra, except for $^{14}$C and tritium, which follow the \texttt{BetaShape}-calculated spectrum~\cite{mougeot_betashape} and the spectral model of Ref.~\cite{katrin:tritium}, respectively. The $^{136}$Xe two-neutrino double-beta decay spectrum follows Ref.~\cite{kotila_phase-space_2012}.]

XELDA has shown that L-shell electron-capture (EC) events of $^{127}$Xe have a smaller charge yield than standard $\beta$ decays~\cite{EC:temples}, suggesting a similar effect for the LM- and LL-shell DEC events of $^{124}$Xe. In \diamx, we model this by introducing shape parameters $Q_{X}/Q_{\beta} \leq 1$ $(X=L, LM, LL)$, defined as the ratio of the charge yield of an $X$-shell EC or DEC event to that of a $\beta$ decay with the same energy. 
This treatment follows the approach adopted by the \LZ collaboration~\cite{LZ:2024zvo}.
As discussed above, the L-shell EC charge yields have been measured by \LZ~\cite{LZ:2025hud}, yielding $Q_L/Q_\beta=0.88$, which implies $Q_{LL}/Q_\beta \leq 0.88$ in the \LZ dataset.
In contrast, the \XEn dataset remains compatible with a model that uses the $\beta$ charge yield~\cite{XENON:2025vwd}, corresponding to $Q_{LL}/Q_\beta = 1$ in the \XEn treatment.

To quantify the impact of uncertainties in the $^{124}$Xe DEC charge yields, we evaluate all candidate DM models under three charge-yield assumptions. 
\textbf{Case I} (baseline): $Q_{LM,LL}/Q_{\beta}=0.88$ for both \LZ and \px, and $1.0$ for \XEn. 
\textbf{Case II}: $Q_{LL}$ for \LZ and \px is treated as a floating parameter. 
\textbf{Case III}: both $Q_{LM}$ and $Q_{LL}$ for \XEn are also floated.
Case I assumes no special treatment of DEC backgrounds and therefore serves as our baseline; unless otherwise stated, all results presented below refer to this case. 
Since \px operates with an electric field strength similar to that of \LZ, we adopt the same charge yield assumptions for \px as for \LZ, rather than the nominal treatment used in the original \px publication~\cite{PandaX:2024qfu}. 
Case II attempts to replicate the DEC background treatment originally adopted by \LZ and \XEn, while Case III represents the most conservative approach, in which all charge-yield parameters not directly measured are treated as unconstrained nuisance parameters.
The results are summarized in Table~\ref{tab:q_ratios}.

An illustrative comparison of the observed events in corrected (S1, S2) space with the expected backgrounds and a representative endothermic inelastic DM benchmark is shown in Fig.~\ref{fig:five_figures}. We define the WIMP search ROI in corrected (S1, S2) space for each dataset the same way as in the corresponding publications: for \LZ\ WS2022 (WS2024), $3 < \mathrm{S1c} < 80~\mathrm{phd}$ and $600~(645) < \mathrm{S2c} < 10^{4.5}~\mathrm{phd}$; for \px, $2 < \mathrm{cS1} < 135~\mathrm{PE}$ and $120 < \mathrm{cS2}_{\text{b}} < 20000~\mathrm{PE}$; for \XEn, $0 < \mathrm{cS1} < 100~\mathrm{PE}$ and $10^{2.1} < \mathrm{cS2} < 10^{4.1}~\mathrm{PE}$. We refer to events with either corrected S1 or S2 above these WIMP-ROI bounds as “outside the ROI.”
\\

\begin{figure}[htpb]
    \centering
    \includegraphics[width=0.75\linewidth]{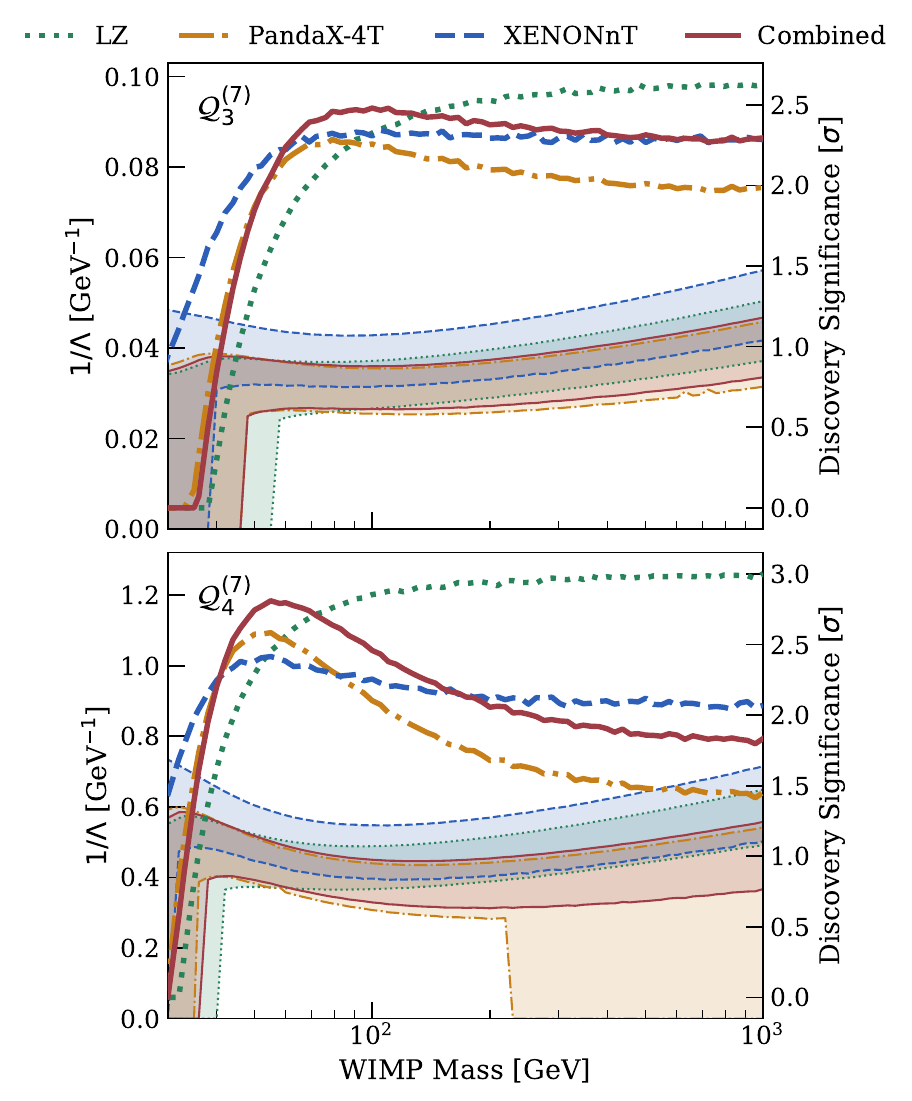}
    \caption{Allowed 90\% C.L. regions for the reciprocal of the cutoff scale $1/\Lambda$ (shaded bands, left $y$-axis) and local discovery significances (lines, right $y$-axis) as functions of the DM mass for the DMEFT operators $\mathcal{Q}_3^{(7)}$ and $\mathcal{Q}_4^{(7)}$. Solid lines denote the combined discovery significances, while dotted, dashdotted and dashed lines correspond to \LZ, \px and \XEn datasets, respectively.
    }
    \label{fig:combined_ChEFT}
\end{figure}

\begin{figure}[ht]
    \centering
    \includegraphics[width=0.95\linewidth]{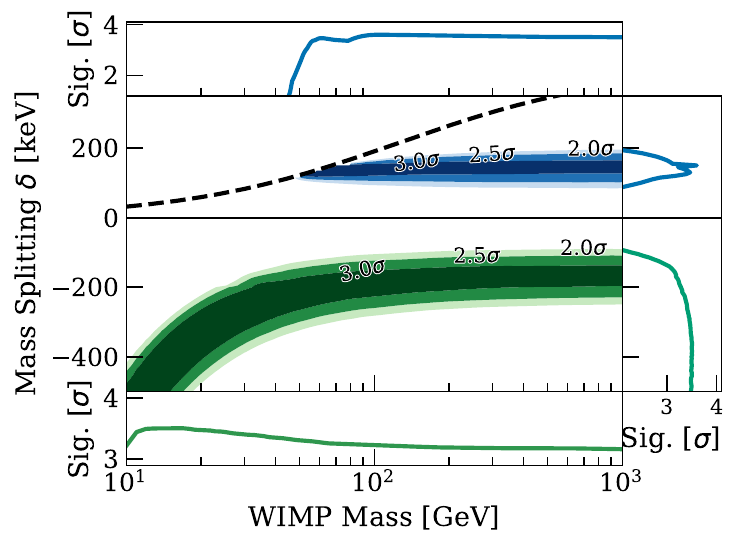}
    \caption{Contours of the local signal significance for endothermic (blue) and exothermic (green) inelastic DM in the $(m_{\mathrm{DM}}, \delta)$ space. The black dashed line marks the kinematic threshold above which endothermic scattering in xenon is kinematically forbidden. The side and top/bottom panels show the corresponding local significances marginalized over $m_{\mathrm{DM}}$ and $\delta$, respectively.}
    \label{fig:combined_inelastic_contours}
\end{figure}

\noindent \textbf{DM interpretation}.
We perform a combined profile-likelihood analysis for each DM model. For each dataset, the likelihood includes event distributions in corrected (S1, S2) space and Gaussian constraints on the background rates; the combined likelihood is the product of the individual dataset likelihoods. Discovery significances and confidence intervals are computed using the asymptotic formulae of Ref.~\cite{cowan_asymptotic_2011}. All quoted significances are local, evaluated at the best-fit points of the corresponding fixed model hypotheses, and do not account for the look-elsewhere effect necessary for identifying an excess. Thus, results should be treated only as goodness-of-fit diagnostics. As a validation, we reproduce the official SI WIMP cross-section upper limits for WIMP masses larger than 50 GeV to within $10\%$ for LZ WS2022, \px Run0+Run1, \XEn SR0, and \XEn SR0+SR1, while for LZ WS2024 our limit is weaker by $\sim 20\%$, which we attribute to the absence of radon-tagging modeling. This level of agreement confirms that our inference procedure is reliable for reinterpreting the high-energy NR-like events.

For the operators $\mathcal{Q}_3^{(7)}$ and $\mathcal{Q}_4^{(7)}$, the Case~I combined-fit results are shown in Fig.~\ref{fig:combined_ChEFT}, with maximal local significances of about $2.3\sigma$ and $2.6\sigma$, respectively. For both operators, the preferred cutoff scale $\Lambda$ lies in the range ${\cal O}(1-10)~{\rm GeV}$, indicating that the interactions are mediated by light states. The significance contours for inelastic DM are shown in Fig.~\ref{fig:combined_inelastic_contours}. In the endothermic case, we find local significances above $3\sigma$ for mass splittings $\delta \simeq 120$–$165~\mathrm{keV}$. In this regime, only DM particles with velocities above the kinematic threshold in Eq.~\ref{eq:vmin} can scatter, naturally concentrating the signal at higher recoil energies while strongly suppressing the rate at lower energies. For exothermic scattering, we obtain local significances up to $3\sigma$ for $\delta < -135~\mathrm{keV}$. The characteristic recoil energy is set by the point where $v_{\min}=0$,
$E_R^{\ast} \equiv |\delta|\frac{m_{\mathrm{DM}}}{m_{\mathrm{DM}} + m_N}$,
which naturally populates the observed high-energy recoil NR-like events while suppressing low-energy events in xenon over the preferred parameter region. The scaling of $E_R^{\ast}$ also explains the $3\sigma$ band in Fig.~\ref{fig:combined_inelastic_contours}: as $m_{\mathrm{DM}}$ decreases, a larger $|\delta|$ is required to keep $E_R^{\ast}$ fixed. Thus, both endothermic and exothermic inelastic scattering can account for the observed high-energy NR-like events without overproducing low-energy events.

\begin{figure}[htpb]
    \centering
    \includegraphics[width=0.9\linewidth]{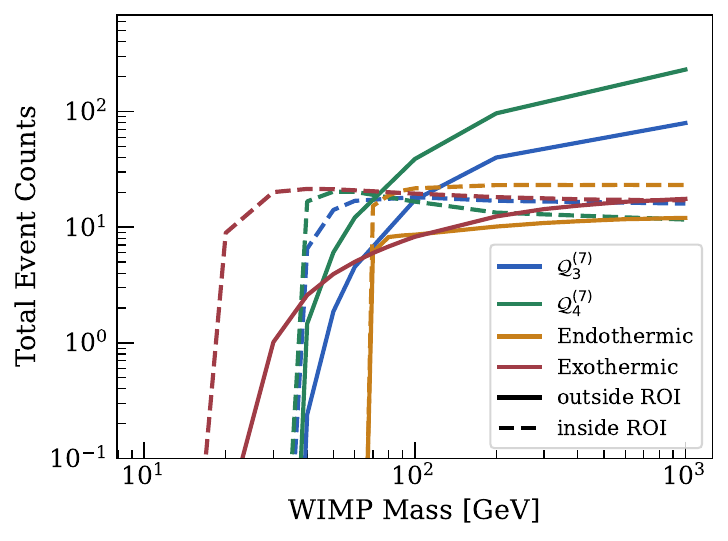}
    \caption{Predicted total number of events inside (dashed) and outside (solid) the WIMP search ROI for the four DM benchmarks, assuming the exposure of the 8.8 \ty of \LZ, \px, and \XEn. The models include $\mathcal{Q}_3^{(7)}$, $\mathcal{Q}_4^{(7)}$, endothermic inelastic scattering with $\delta = 150~\mathrm{keV}$, and exothermic inelastic scattering with $\delta = -200~\mathrm{keV}$. All four are evaluated at their 3-experiment combined best-fit normalizations. }
    \label{fig:events-in-high-Er}
\end{figure}

Using a set of representative benchmarks, we also predict the event rate at higher NR energies. In Fig.~\ref{fig:events-in-high-Er}, we show four DM benchmarks as functions of $m_{\mathrm{DM}}$, with the predicted numbers of events inside (dashed) and outside (solid) the ROI. For the two velocity-dependent benchmarks, the out-of-ROI yield exceeds $\mathcal{O}(30)$ events at large $m_{\mathrm{DM}}$.  
The event yield from $\mathcal{Q}_4^{(7)}$ is significantly larger than that from $\mathcal{Q}_3^{(7)}$, owing to their different $q$-scaling.
Since the \LZ and \px data at higher NR energy show no corresponding large excess under a nominal background model~\cite{LZ:2023lvz,LZ:2024vge,PandaX:2024qfu}, we infer $m_{\mathrm{DM}} \lesssim 110~\mathrm{GeV}$ and $70~\mathrm{GeV}$ for $\mathcal{Q}_3^{(7)}$ and $\mathcal{Q}_4^{(7)}$, respectively.
Complementary information is provided by Fig.~\ref{fig:nr_below_median}, which shows the NR spectra of out-of-ROI events for the $\mathcal{Q}_3^{(7)}$ and exothermic/endothermic benchmarks. The spectra peak in the range $50$–$200~\mathrm{keV}$, indicating where the predicted out-of-ROI events are concentrated. 

For inelastic DM, we take $\delta = -200$ and $150~\mathrm{keV}$ for exothermic and endothermic scattering. In the endothermic case, the out-of-ROI rate turns on around $m_{\mathrm{DM}} \sim 70~\mathrm{GeV}$ sharply and saturates at $\sim 10$ events for $m_{\mathrm{DM}} \gtrsim 500~\mathrm{GeV}$, while in the exothermic case it grows to $\sim 30$ events at TeV-scale masses and then levels off. In both inelastic benchmarks, the predicted high-recoil yield is at the level of a few tens of events, providing a clear target for upcoming analysis of  \LZ, \px and \XEn data.

Finally, we assess the robustness of these conclusions under different DEC charge-yield assumptions. As shown in Table~\ref{tab:q_ratios}, the formal local significances for our benchmark DM models are sensitive to this choice: they can exceed $3.5\sigma$ in our baseline assumption (Case I),
but fall below $1.1\sigma$ for all models in the most conservative case (Case III). This spread shows that current DEC systematics dominate the quantitative significance, even though the qualitative preference for spectra that enhance high energy NR events remains. \\

\noindent \textbf{Discussions and Conclusions}.~~~We investigate DM interpretations of the high-energy NR-like events in recent liquid-xenon experiments. Using our unified \diamx framework, built on publicly available data and likelihood models, we perform the first combined profile-likelihood fits across multiple WIMP-search datasets, totaling 8.8~\ty of exposure. We show that two broad classes of DM–nucleon interactions—velocity-dependent cross sections and inelastic scattering (both endo- and exothermic)—can reproduce the observed high-energy NR-like events, corresponding to local statistical significances of up to $3.5\sigma$. We quantify the impact of the $^{124}$Xe DEC background, finding that plausible variations in the poorly known DEC charge yield can shift the inferred local significance for DM models down to below $1\sigma$. 

Within DMEFT, the velocity-dependent operators $\mathcal{Q}_3^{(7)}$ and $\mathcal{Q}_4^{(7)}$ can accommodate these events at the few–$\sigma$ level. The best-fit values imply cutoff scales of ${\cal O}(1-10)$ GeV as shown in Fig.~\ref{fig:combined_ChEFT}. When combined with collider searches~\cite{Goodman:2010ku,An:2012va}, this suggests that if these interactions are responsible for these NR-like events, the DM-gluon coupling must be mediated by light states. For endothermic and exothermic inelastic DM scenarios, the cutoff scale $\Lambda$ is ${\cal O}(10)$ TeV (see Table.~\ref{tab:q_ratios}), safely evading collider bounds.

Using the best-fit benchmarks, we also predict the event rate at higher NR energies outside the present ROI. Velocity-dependent models tend to overproduce such events at large $m_{\mathrm{DM}}$, while inelastic benchmarks robustly forecast only a few tens of high-recoil events. Improved measurements of DEC charge yields, together with extended high-recoil searches in upcoming analysis of \LZ, \px and \XEn data, will therefore be crucial to determining whether these events arise from DM or from incompletely modeled backgrounds.

~\\
\noindent \textbf{Acknowledgments}.
We thank the \LZ, \px, and \XEn collaborations for their publicly available results. We are particularly grateful to Yi Tao for assistance in analyzing the PandaX-4T data.
The work of H.A., H.N. and C.X. is supported in part by the National Key R\&D Program of China under Grant Nos. 2023YFA1607104 and 2021YFC2203100 and the National Science Foundation of China under Grant Nos. 12475107 and 12525506. 
The work of F.G. is supported by the National Science Foundation of China under Grant No. 12521007 and the Ministry of Education of China under Grant No. SRICSPYF-ZY2025028. 
H.A. and F.G. acknowledge support from the Dushi program of Tsinghua University and the Tsinghua University Initiative Scientific Research Program.
The work of J.L. is supported by the National Science Foundation of China under Grant Nos. 12235001 and 12475103, and State Key Laboratory of Nuclear Physics and Technology under Grant No. NPT2025ZX11.

\bibliographystyle{elsarticle-num}
\bibliography{axion-rev}

\end{document}